\documentclass[a4paper,11pt,fleqn]{article}
\usepackage{amsfonts,latexsym,epsfig}

\def\nn{\nonumber}
\def\non{\nonumber\\}
\def\be{\begin{equation}}
\def\ee{\end{equation}}
\def\ben{\begin{displaymath}}
\def\een{\end{displaymath}}
\def\ba{\begin{eqnarray}}
\def\ea{\end{eqnarray}}

\def\a{\alpha}

\def\d{\delta}

\def\G{\Gamma}

\def\m\mu 
\def\n{\nu}
\def\Om{\Omega}

\def\r{\rho}
\def\s{\sigma}

\def\t{\tau}

\def\moth{\mathsurround=0pt}
\newdimen\zo \zo=0pt

\def\tick{\leaders\hrule height 0.5ex depth 0pt \hskip 0.5pt}
\def\upboxfill{$\moth \setbox\zo\hbox{\tick}%
  \hskip 2pt\hbox to 0pt{$\tick$\hss}\hrulefill \hbox to 6pt{$\tick$\hss}$}

\def\dtick{\leaders\hrule height .34pt depth .5ex \hskip 0.5pt}
\def\downboxfill{$\moth \setbox\zo\hbox{\dtick}%
  \hskip 2pt\hbox to 0pt{$\dtick$\hss}\hrulefill \hbox to 6pt{$\dtick$\hss}$}
\def\overbox#1{\mathop{\vbox{\moth\ialign{##\crcr\noalign{}
\downboxfill\crcr\noalign{\vskip 1pt\nointerlineskip}
      $\hfil\displaystyle{#1}\hfil$\crcr}}}\limits}

\def\oversym#1{\overbox{{}#1}}

\def\cI{{\cal I}}

\def\cM{{\cal M}}
\def\cO{{\cal O}}
\def\cP{{\cal P}}
\def\cR{{\cal R}}

\def\cV{{\cal V}}
\def\cX{{\cal X}}
\def\cY{{\cal Y}}

\def\rt{{\tilde{\r}}}

\def\Pt{{\tilde{\Pi}}}

\def\Z{\mathbb{Z}}

\def\N{\mathbb{N}}

\def\E{E_8}
\def\eee{{\mathfrak{e}_8}}
\def\eeee{{\mathfrak{e}_9}}
\def\EE{E_{8(8)}}
\def\9{E_{9(9)}}
\def\0{E_{10}}
\def\SO{{SO}(16)}
\def\so{{\mathfrak{so}(16)}}

\def\la{\label}
\def\ci{\cite}

\def\Ref#1{(\ref{#1})}

\def\ft#1#2{{\textstyle {\frac{#1}{#2}} }}
\def\8{\infty}

\def\tr{{\rm tr \,}}
\def\ra{\rightarrow}
\def\i{{\rm i}\hbar}

\def\16{N\!=\! 16}
\def\216{d\!=\!2 , N\!=\!16}
\def\316{d\!=\!3 , N\!=\!16}
\def\48{d\!=\!4 , N\!=\!8}

\def\gg {\mathfrak{g}}

\def\gk{\mathfrak{k}}

\makeatletter
\@addtoreset{equation}{section}
\makeatother

\newcommand{\ID}{\mbox{1\hspace{-.35em}1}}

\begin{document}

\thispagestyle{empty}

\begin{flushright}
AEI-103\\
LPTENS-99/08\\
hep-th/9903111
\end{flushright}
\renewcommand{\thefootnote}{\fnsymbol{footnote}}

\vspace*{0.1cm}
\begin{center}
\mathversion{bold}
{\bf\Large  On the Yangian [$Y\!(\eee)$] quantum symmetry of}\medskip

{\bf\Large maximal supergravity in two dimensions}\bigskip\bigskip
\mathversion{normal}

{\bf\large K.~Koepsell, H.~Nicolai\medskip\\ }
{\large Max-Planck-Institut f{\"u}r Gravitationsphysik,\\
Albert-Einstein-Institut,\\
Schlaatzweg 1, D-14473 Potsdam, Germany}~\footnote{Research supported
in part by EC under TMR contract ERBFMRX-CT96-0012.}$^,$\footnote{New
address after April 1, 1999: Haus 5, Am M{\"u}hlenberg, 14476 Golm, Germany.}
\smallskip\\ {\small E-mail:
koepsell@aei-potsdam.mpg.de, nicolai@aei-potsdam.mpg.de} 
\bigskip

\setcounter{footnote}{0}
{\bf\large H.~Samtleben\medskip\\ }
{\large Laboratoire de Physique Th{\'e}orique\\
de l'Ecole Normale Sup{\'e}rieure,\\
24 Rue Lhomond, 75231 Paris Cedex 05, France}~\footnotemark$^,\!$
\setcounter{footnote}{2}\footnote{Unit{\'e} Mixte de Recherche du Centre
National de la Recherche Scientifique et de l'Ecole Normale Sup{\'e}rieure.} 
\smallskip\\ {\small E-mail:
henning@lpt.ens.fr\medskip} 
\end{center}
\renewcommand{\thefootnote}{\arabic{footnote}}
\setcounter{footnote}{0}
\bigskip
\medskip
\begin{abstract}
We present the algebraic framework for the quantization of the
classical bosonic charge algebra of maximally extended ($N\!=\!16$)
supergravity in two dimensions, thereby taking the first steps towards
an exact quantization of this model. At the core of our construction
is the Yangian algebra $Y(\eee)$ whose $RTT$ presentation we discuss
in detail.  The full symmetry algebra is a centrally extended twisted
version of the Yangian double ${\cal D}Y(\eee)_c$. We show that there
exists only one special value of the central charge for which the
quantum algebra admits an ideal by which the algebra can be divided so
as to consistently reproduce the classical coset structure $\EE/\SO$
in the limit $\hbar\!\ra\!0$.
\end{abstract}

\renewcommand{\thefootnote}{\arabic{footnote}}
\setcounter{footnote}{0}

\newpage

\section{Introduction} 
Dimensionally reduced gravity and supergravity are well known to
possess hidden symmetries \cite{CreJul79,Juli80}. Of special interest
in this context is the case of two dimensions, where these symmetries
become infinite dimensional, generalizing the so-called Geroch group
of general relativity \cite{Juli81,BreMai87,JulNic96}.  The existence
of infinite dimensional symmetries in these models is intimately
linked to their integrability, which is borne out by the existence of
linear systems for their classical equations of motion, both for the
bosonic models \cite{Mais78,BelZak78,BreMai87} and their locally
supersymmetric extensions \cite{Nico87,NicWar89}. In this paper, we
will focus attention on the maximally extended $N\!=\!16$
supergravity, whose scalar sector is governed by an $\EE/\SO$
nonlinear $\sigma$-model, and whose equations of motion admit a rigid
non-compact $\9$ symmetry.
\footnote{This symmetry acts as a solution generating ``isometry
group'', or as a group of ``dressing transformations''
\cite{BelZak78,Seme85,BabBer92,BerJul97}.}

The canonical structure of these models and the Lie-Poisson
realization of the associated infinite dimensional symmetries were
analyzed only quite recently \cite{KorSam98,NicSam98}. As shown there,
the affine Lie algebra seen at the level of the classical equations of
motion is converted into a quadratic algebra of Yangian type in the
canonical formulation. One key feature of this result, which we
exploit in this paper, is that the quadratic algebra, and therefore at
least part of the model, can be quantized directly by replacing the
Poisson algebra of charges by an exchange algebra involving a suitable
$R$ matrix, whereas a standard field theoretic quantization would
appear to be prohibitively difficult. The relevant $R$-matrix based on
the exceptional group $\E$ has already been derived in
\cite{ChaPre91a}.  The structure that appears upon quantization is the
Yangian $Y(\eee)$. As a consequence, the physical states of the
quantized theory must belong to multiplets of $Y(\eee)$ rather than
multiplets of the affine algebra $\eeee$ as one might have naively
expected.

The Yangian of the exceptional algebra $\eee$ is distinguished from
the Yangians of the classical Lie algebras by the fact that its
fundamental representation is reducible over $\eee$, namely decomposes
into ${\bf 249}={\bf 1}\!\oplus\!{\bf 248}$. The $R$-matrix associated
to this representation has been given by Chari and Pressley in
\cite{ChaPre91a}. Using their result and the general analysis of
Drinfeld \cite{Drin85} we obtain the $RTT$ presentation of $Y(\eee)$
which may be viewed \cite{FaReTa90} as the quantization of
group-valued $E_8$ matrices endowed with the symplectic structure of
dimensionally reduced gravity. The full quantum structure that appears
upon quantization of the algebra of classical nonlocal charges is a
centrally extended twisted version of the Yangian double, that
reflects the $\EE/\SO$ coset structure of the classical model.

The presence of this extra coset structure and its quantum consistency
require further properties of the $Y(\eee)$ $R$-matrix beyond those
discussed in \cite{ChaPre91a}. We explain these in detail here. In
particular, for a discrete set of values of the central extension, the
algebra ${\cal D}Y(\eee)_c$ possesses nontrivial ideals which may be
divided out to reduce the number of degrees of freedom.  Remarkably
there is only one among the altogether eight ``exceptional'' values of
the central extension, which admits a non-trivial ideal for which the
quantum monodromy matrix becomes symmetric in the limit
$\hbar\!\ra\!0$ and the associated ideal can be consistently divided
out to recover the classical coset space $\E/\SO$ of $N\!=\!16$
supergravity. The relevant value of the central extension ($c\!=\!1$
with our normalization) differs from the critical value $c\!=\!15$ for
which the quantum algebra admits an additional infinite-dimensional
center \cite{ResSem90}.

The main open problem which remains is the compatibility of the local
supersymmetry constraints with the $Y(\eee)$ charge algebra at the
quantum level. In this paper we have concentrated on the direct
quantization of the algebra of nonlocal charges which are classically
invariant under supersymmetry, i.e.~Poisson commute weakly with the
supersymmetry generators. A complete treatment should in addition
contain a quantum version of the supersymmetry constraint algebra (an
$N\!=\!16$ superconformal algebra) which could serve to define the
physical states as its kernel. The Yangian structure exhibited in this
paper would then become a spectrum generating algebra for $N\!=\!16$
supergravity. Let us emphasize, however, that the interplay between
canonical constraints and non-local conserved charges in integrable
field theories has so far not been studied at all at the quantum
level, as the existing literature deals exclusively with flat space
models rather than the generally covariant and locally supersymmetric
models we are concerned with here.

Our results underline the importance of quantum group structures for
dimensionally reduced gravity and supergravity. The ultimate aim here
is the identification of a ``quantum Geroch group'' which would act on
the space of physical states in the same way as the classical Geroch
group acts on the moduli space of classical solutions. The relevance
of these structures for string and $M$-theory seems also obvious.
After all, the resulting symmetries can be regarded as quantum
deformations of the infinite dimensional $U$-duality symmetries that
have been conjectured to appear in compactified string and $M$-theory
\cite{HulTow95}.  However, our results also indicate that some widely
held perceptions and expectations may need to be revised. In
particular, the underlying symmetry of the full quantum theory may
turn out to be related to some (hyperbolic?) extension of $Y(\eee)$,
rather than just the arithmetic duality groups $\9 ({\Z})$ and
$E_{10(10)}(\Z)$.

\mathversion{bold}
\section{$E_8$ preliminaries}
\mathversion{normal} In this section we collect some basic facts on
the exceptional algebra $\eee$ thereby fixing the notation for the
following. In particular, we give very explicit expressions for the
projectors onto the irreducible parts of the tensor product of two
adjoint representations of $\eee$.

The generators of $\eee$ in the adjoint (and thus fundamental)
representation are denoted by $X^a$. We are here interested its
non-compact maximally split form with maximal compact subalgebra
$\mathfrak{so}(16)$, giving rise to the coset space $E_{8(8)}/SO(16)$.
Accordingly, we split $\E$ indices $a,b, \dots$ as $([IJ],A),\dots $,
with $I,\,J=1,\ldots,16$ and $A=1,\ldots,128$ corresponding to the
decomposition $\mbox{\bf 248}\ra\mbox{\bf 120}\oplus\mbox{\bf 128}$ of
the adjoint representation of $\eee$ into the adjoint and the
fundamental spinor representation of $\mathfrak{so}(16)$.  The
generators satisfy the commutation relations
\be
  \big[X^a,X^b\big] = f^{ab}{}_c\,X^c\;,
\ee
with the convention that summation over antisymmetrized pairs of 
indices $[IJ]$ is always accompanied by a factor $\frac12$, viz.
\ben
X^a Y_a \equiv X^A Y_A + \ft12 X^{IJ} Y_{IJ}\;.
\een
The structure constants are most conveniently given in their
fully antisymmetric form obtained by raising the index $c$ with 
the help of the Cartan-Killing form $\eta^{ab}$. For the adjoint 
representation, the latter is defined by 
\be
\eta^{ab} := \ft1{60} \tr\big(X^{a}X^{b}\big)
           = \ft1{60} f^{ac}{}_d f^{bd}{}_c
\ee
which yields 
\be
\eta^{AB}=\d^{AB}\;, \qquad
\eta^{I\!J\,K\!L}=-2\,\d^{IJ}_{KL}\;.
\ee
The $\eee$ structure constants are then completely characterized by
\be
f^{I\!J,\,K\!L,\,M\!N} = 
-8\, \d\!\oversym{^{I[K}_{\vphantom{M}}\,\d_{MN}^{L]J}},\qquad
f^{I\!J,\,A,\,B}   = -\ft12 \G^{IJ}_{AB}, \la{fabc}
\ee
where the matrices $\G^{IJ}_{AB}$ are obtained from the
$\mathfrak{so}(16)$ $\G$-matrices in the standard fashion
\be
\G^{I}_{A\dot{A}}\G^{J}_{\dot{A}B}=\d^{IJ}_{AB} +\G^{IJ}_{AB}.
\ee
The maximal compact subalgebra $\mathfrak{so}(16)$ can be
characterized alternatively as the subalgebra invariant under 
the symmetric space involution 
\be
\t (X^a) = - (X^a)^T\;.
\ee

For the formulation of the Yang Baxter equation we will need
to deal with operators acting acting on the tensor product 
${\bf 248}\!\otimes\!{\bf 248}$. The associated matrices will be
denoted as $O_{ab}{}^{cd}$, where we refer to the indices $ab$ as 
``incoming'' and to the indices $cd$ as ``outgoing''. The product 
of two such matrices $O$ and $P$ is consequently given by
\ben
(OP)_{ab}{}^{cd}:=O_{ab}{}^{ef} P_{ef}{}^{cd}\;.
\een
As with the generators above, the Cartan-Killing metric must be
used whenever indices are raised or lowered from their
``canonical'' position on such matrices. We define
\be
\stackrel{21}{O}_{ab}{}^{cd} ~:=~ \stackrel{12}{O}_{ba}{}^{dc}\;.
\ee
To write down the projectors we need the operators $\ID$, $\Pi$ (i.e. 
the identity and the exchange operator, respectively), and $\Pt$,
which are given by
\be
\ID_{ab}{}^{cd}   = \d^c_a\,\d^d_b, \quad
\Pi_{ab}{}^{cd} = \d^d_a\,\d^c_b, \quad
\Pt_{ab}{}^{cd} = \eta_{ab}\,\eta^{cd}. 
\ee
A further important operator is the symmetric Casimir element 
defined in the adjoint representation by
\ba
\Om_\eee&\equiv& \eta{}_{ab}\; X^a\!\otimes\!X^b \\
& = & 
   - \ft1{2}\, X^{IJ}\!\otimes\!X^{IJ} 
   + X^A\!\otimes\!X^A ~\in~ \so\otimes\so + \gk\otimes\gk \;. \nn
\ea
In terms of the structure constants of $\E$, the Casimir element 
can be alternatively expressed as
\be\la{O1}
(\Om_\eee)_{ab}{}^{cd} = f^e{}_a{}^c f_{eb}{}^d \;.\nn
\ee
We will also need the twisted Casimir element $\Om_\eee^\t$,
defined by 
\ba
\Om_\eee^\t &\equiv& \eta_{ab} X^a \otimes \t (X^b) \\
   &=&  - \ft1{2}\, X^{IJ}\!\otimes\!X^{IJ} 
        - X^A\!\otimes\!X^A \;,\nn
\ea
i.e.~in indices:
\be\la{O2}
(\Om_\eee^\t)_{ab}{}^{cd} = - (\Om_\eee)_{cb}{}^{ad} =
     - f_{ea}{}^c f_{eb}{}^d \;.
\ee
The twisted Casimir element is obviously not $\E$ but only $\SO$
        invariant. 

The tensor product of two adjoint representations of
$\E$ splits into its irreducible components according to 
${\bf 248}\!\otimes\!{\bf 248} = {\bf 1}\!\oplus\!{\bf 248}\!\oplus\! 
{\bf 3875}\!\oplus\!{\bf 27000}\!\oplus\!{\bf 30380}$. 
The corresponding projectors are given by:
\ba
\cP_1 &=& \ft1{248}\,\tilde\Pi,\la{proj}\\
\cP_{248} &=& \ft1{60}\left(\Om_\eee\Pi-\Om_\eee\right),\non
\cP_{3875} &=& \ft1{14}\,\left(\ID-\ft14\,\tilde\Pi+\Pi
              -\ft12(\Om_\eee\Pi+\Om_\eee)\right),\non
\cP_{27000} &=& \ft17\left(3\,\ID +\ft3{31}\,\tilde\Pi+3\,\Pi
              +\ft14(\Om_\eee\Pi+\Om_\eee)\right),\non
\cP_{30380} &=& \ft12\,\ID -\ft12\,\Pi
              +\ft1{60}\left(\Om_\eee-\Om_\eee\Pi\right).\nn
\ea
To verify that these operators indeed satisfy orthogonal projection 
relations, one needs the following relation
\be
\Om_\eee^2 = 12\, \ID +12\,\Pi +12\,\Pt -20\,\Om_\eee +10\,\Om_\eee\Pi\;,
\ee
whose validity we have established with the help of a computer.
In terms of the $\eee$ structure constants this relation becomes
\ben
f^e{}_{ag}f_{beh}f^{gic}f_i{}^{hd} =
 24 \d_{(a}^{\hphantom{(}c} \d_{b)}^d +12\eta_{ab}\eta^{cd}
-20 f^e{}_a{}^c f_{eb}{}^d +10f^e{}_a{}^d f_{eb}{}^c\;. 
\een
In indices, the projectors read:
\ba 
(\cP_{1})_{ab}{}^{cd}     &=& \ft{1}{248}\,\eta_{ab} \eta^{cd},\la{projind}\\
(\cP_{248})_{ab}{}^{cd}   &=& -\ft{1}{60}\, f^e{}_{ab} f_e{}^{cd},\non
(\cP_{3875})_{ab}{}^{cd}  &=& \ft{1}{7}\,  \d_{(a}^{\hphantom{(}c} \d_{b)}^{d} 
                             -\ft{1}{56}\, \eta_{ab} \eta^{cd}
                             -\ft{1}{14}\, f^e{}_a{}^{(c} f_{eb}{}^{d)},\non
(\cP_{27000})_{ab}{}^{cd} &=& \ft{6}{7}\, \d_{(a}^{\hphantom{(}c} \d_{b)}^{d} 
                             +\ft{3}{217}\,\eta_{ab} \eta^{cd}
                             +\ft{1}{14}\, f^e{}_a{}^{(c} f_{eb}{}^{d)},\non
(\cP_{30380})_{ab}{}^{cd} &=& \d_{[a}^c \d_{b]}^d
                             +\ft{1}{60}\, f^e{}_{ab} f_e{}^{cd}.\nn
\ea
All these projectors are manifestly symmetric w.r.t. interchange of
the two subspaces, i.e. $\stackrel{12}\cP_j~=~\stackrel{21}\cP_j$.
Furthermore, any $E_8$ matrix $\cV$ obeys 
\be
\cP_j \, \cV\otimes \cV = \cV\otimes\cV \, \cP_j \;,
\ee
which together with the normalization $\det\cV=1$ can be taken as
defining relations for the group elements of $E_8$.

\mathversion{bold}
\section{$R$-matrix and the Yangian $Y(\eee)$}
\mathversion{normal}

Here, we review the Yangian algebra $Y(\eee)$ and the $R$-matrix
associated to its fundamental representation ${\bf 249}$
\cite{ChaPre91a}. The Yangian $Y(\eee)$ \ci{Drin85} is recursively
defined as the associative algebra with generators $\cX^a$ and $\cY^a$
$(a=1,\ldots,248)$ and relations
\ba
&&\left[\cX^a,\cX^b\right] ~=~ \i f^{ab}{}_c\,\cX^c\;,\qquad
\left[\cX^a,\cY^b\right] ~=~ \i f^{ab}{}_c\,\cY^c\;,\la{yangian}\\[4pt]
&&\left[\cY^a\left[\cY^b,\cX^c\right]\right] -
\left[\cX^a\left[\cY^b,\cY^c\right]\right] ~=~ -\hbar^2 L^{abc}\;,\non[6pt]
&&\mbox{with~~}
L^{abc}=\ft{1}{24}\,f^{ad}{}_g f^{be}{}_h f^{cf}{}_i f^{ghi}
\{\cX^d, \cX^e, \cX^f\}\;,\non
&&\mbox{and~~}
\{\cX^1, \cX^2, \cX^3\}=\sum_\s \cX^{\s(1)}\cX^{\s(2)}\cX^{\s(3)}\;.\nn
\ea
It admits a nontrivial coproduct and antipode structure whose 
explicit form will not be needed here, see e.g.~Thm. 12.1.1
of \cite{ChaPre94} for details.

Due to the fact that $L^{abc}$ does not vanish when the $\cX^a$ are
evaluated in the fundamental representation of $\eee$, it is not
possible to lift this representation of $\eee$ to a representation of
$Y(\eee)$. Rather, the minimal representation of $Y(\eee)$ is
reducible over $\eee$ and contains an additional trivial
representation of $\eee$ \ci{Drin85}. With respect to $\so$ we thus
have the decomposition
\be
\mbox{\bf 249}\ra\mbox{\bf 1}\oplus\mbox{\bf 120}\oplus\mbox{\bf 128}
\;.\la{sodec}
\ee
For compactness of notation, we will label the extra singlet by $0$ 
and use hatted indices which run over all 249 dimensions, i.e.
$0\le \hat{a}, \hat{b},\dots \le 248$.

The $R$-matrix associated with the fundamental representation of
$Y(\eee)$ is the solution $R(w)$ to the Quantum Yang-Baxter Equation
($\equiv$ QYBE)  
\be
\stackrel{12}R (u-v) \stackrel{13}R (u) \stackrel{23}R (v) =
\stackrel{23}R (v) \stackrel{13}R (u) \stackrel{12}R (u-v) \;,
\ee
or, with indices written out,
\be\la{QYB}
R_{\hat{a}\hat{b}}{}^{\hat{g}\hat{h}}(u\!-\!v)
R_{\hat{g}\hat{c}}{}^{\hat{p}\hat{i}}(u)
R_{\hat{h}\hat{i}}{}^{\hat{q}\hat{r}}(v) ~=~
R_{\hat{b}\hat{c}}{}^{\hat{h}\hat{i}}(v)
R_{\hat{a}\hat{i}}{}^{\hat{g}\hat{r}}(u)
R_{\hat{g}\hat{h}}{}^{\hat{p}\hat{q}}(u\!-\!v)\;,
\ee
The classical limit is 
\be\la{Rcl}
R(w) = \ID - \frac{\i}{w}\;\Om_{\eee} 
            + \cO\left(\frac{\hbar^2}{w^2}\right) 
\qquad \mbox{for}\quad w \ra \8 \;.
\ee
where the definition of the Casimir element $\Om_\eee$ is extended to
${\bf 1}\oplus{\bf 248}$ by the trivial (zero) action on the ${\bf 1}$.
We also impose the standard normalization condition
\be
R(0) = \Pi\;.
\ee
Within the tensor product ${\bf 249}\!\otimes\!{\bf 249}$ we introduce
in addition to the operators from \Ref{proj} the projector $\cP_{0}$
onto the one-dimensional space ${\bf 1}\!\otimes\!{\bf 1}$ and the
projectors $\cP_{+}$ and $\cP_{-}$ onto the symmetric and
antisymmetric part of the space $({\bf 248}\!\otimes\!{\bf
1})\oplus({\bf 1}\!\otimes\!{\bf 248})$, respectively.  Furthermore,
there are $\eee$ invariant intertwining operators between 
subspaces of the same dimension, which we denote by $\cI_{01}$,
$\cI_{10}$, $\cI_{+248}$, and $\cI_{248+}$. They are defined by 
\ba
\cI_{01} \cI_{10} &=& \cP_{0} \qquad \cI_{10} \cI_{01}=\cP_{1} \;,\\
\cI_{+248}\cI_{248+} &=& \cP_{+} \qquad \cI_{248+}\cI_{+248}=\cP_{248}
\;,\nn
\ea
respectively, up to relative factors between the intertwiners
which drop out in the above relations. Explicitly, the new 
projectors and intertwiners are given by 
\ba
(\cP_{0})_{00}{}^{00}    &=& 1,\la{proj2}\\
(\cP_{+})_{a0}{}^{b0}    &=& (\cP_{+})_{a0}{}^{0b}~=~(\cP_{+})_{0a}{}^{b0}
                          ~=~(\cP_{+})_{0a}{}^{0b}~=~ \ft12\d^a_b,\non
(\cP_{-})_{a0}{}^{b0}    &=&  -(\cP_{-})_{a0}{}^{0b}~=~-(\cP_{-})_{0a}{}^{b0}
                          ~=~(\cP_{-})_{0a}{}^{0b}~=~ \ft12\d^a_b,\non[3ex]
(\cI_{01})_{00}{}^{ab}   &=& \eta^{ab},\non 
(\cI_{10})_{ab}{}^{00}   &=& \ft{1}{248}\eta_{ab},\non
(\cI_{+248})_{0a}{}^{bc} &=& (\cI_{+248})_{a0}{}^{bc} 
                          ~=~ \ft{1}{120} f_a{}^{bc},\non
(\cI_{248+})_{ab}{}^{c0} &=& (\cI_{248+})_{ab}{}^{0c} ~=~ -f_{ab}{}^{c},\nn
\ea
with all other components vanishing. Again all operators are symmetric 
w.r.t.~interchange of the two subspaces with the exception of the 
intertwiners $\cI_{+248}$ and $\cI_{248+}$, which obey
\be
\stackrel{12}\cI_{+248} = - \stackrel{21}\cI_{+248}\;, \qquad
\stackrel{12}\cI_{248+} = -\stackrel{21}\cI_{248+}\;.
\ee
\bigskip

As shown in \ci{ChaPre91a}, the $R$-matrix associated to the fundamental 
representation of $Y(\eee)$ in terms of these projectors and intertwiners 
is given by
\ba
f^{-1}(w)R(w) &=& \ft{w+\i}{w-\i}\;\cP_{30380} +\cP_{27000} 
 +\ft{w^3+15w^2\i+44w(\i)^2+60(\i)^3}{(w-\i)(w-6\i)(w-10\i)}\;\cP_{248}\non
&& {} +\ft{(w+\i)(w+6\i)}{(w-\i)(w-6\i)}\;\cP_{3875}
 +\ft{w^3-15w^2\i+44w(\i)^2-60(\i)^3}{(w-\i)(w-6\i)(w-10\i)}\;\cP_{+}\non
&& {} +\ft{w+\i}{w-\i}\;\cP_{-} 
 +\ft{w^4+30w^3\i+269w^2(\i)^2+660w(\i)^3+900(\i)^4}{
  (w-\i)(w-6\i)(w-10\i)(w-15\i)}\;\cP_{1}\non
&& {}+\ft{w^4-30w^3\i+269w^2(\i)^2-660w(\i)^3+900(\i)^3}{
  (w-\i)(w-6\i)(w-10\i)(w-15\i)}\;\cP_{0}\non
&& {} +\ft{w(\i)^3}{\a^2(w-\i)(w-6\i)(w-10\i)(w-15\i)}\;\cI_{01}\non
&& {} +\ft{248 (60\a)^2 w(\i)^3}{(w-\i)(w-6\i)(w-10\i)(w-15\i)}\;\cI_{10}\non
&& {} -\ft{60\sqrt{2} w(\i)^2}{\a(w-\i)(w-6\i)(w-10\i)}\;\cI_{+248}\non
&& {} +\ft{30\sqrt{2} \a w(\i)^2}{(w-\i)(w-6\i)(w-10\i)}\;\cI_{248+}\;.\nn
\ea
We rewrite this in the form
\ba
f^{-1}(w)R(w) &=& \ID +\sum_{j=1}^4  \frac{\cR_j}{w- w_j}\;, \la{R}
\ea
where the poles are located at
\be
w_1 = \i \: , \; w_2 = 6\i \: , \; w_3 = 10\i \; , \; w_4 = 15\i\;,
\ee
and the associated residues are
\ba
\cR_1 &=& 
      2\,\cP_{30380}-\ft{14}{5}\;
         \cP_{3875}+\ft83\;\cP_{248}-\ft23\;\cP_{+}+2\,\cP_{-}
      -\ft{62}{21}\;\cP_{1}\non
&&   {} -\ft{16}{21}\;\cP_{0}
      -\ft{4\sqrt{2}}{3\a}\;\cI_{+248}+\ft{2\sqrt{2}\a}{3}\;\cI_{248+}
      -\ft{9920\a^2}{7}\;\cI_{10}-\ft{1}{630\a^2}\;\cI_{01}\;, \non
\cR_2 &=& \ft{84}{5}\;\cP_{3875}-54\,\cP_{248}+6\,\cP_{+}
      +124\,\cP_{1}+8\,\cP_{0}+\ft{18\sqrt{2}}{\a}\;\cI_{+248}\non
&& {} - 9\sqrt{2}\a \;\cI_{248+}
      +29760\a^2\,\cI_{10}+\ft{1}{30\a^2}\;\cI_{01}\;,
      \non[8pt]
\cR_3 &=& \ft{250}{3}\;\cP_{248}-\ft{10}{3}\;\cP_{+}
      -\ft{1240}{3}\;\cP_{1}-\ft{20}{3}\;\cP_{0}-
      \ft{50\sqrt{2}}{3\a}\;\cI_{+248}\non
&& {} +\ft{25\sqrt{2}\a}{3}\;\cI_{248+}
      -49600\a^2\,\cI_{10}-\ft{1}{18\a^2}\;\cI_{01}\;,
      \non[8pt]
\cR_4 &=& 
      \ft{2480}{7}\;\cP_{1}+\ft{10}{7}\;\cP_{0}
      +\ft{148800\a^2}{7}\;\cI_{10}+\ft{1}{42\a^2}\;\cI_{01}\;.  \la{resi}
\ea

\noindent
The scalar function $f$ is uniquely defined by its functional
equation 
\be\la{f1}
f(w)f(w-15\i) = \frac{(w-\i)(w-6\i)(w-10\i)(w-15\i)}
{w(w-5\i)(w-9\i)(w-14\i)}\;,
\ee
and its asymptotic behavior
\be\la{f2}
f(w) = 1 - \ft{2\i}{w} + \cO\left(\ft1{w^2}\right)\quad
\mbox{for} \quad w \ra \pm \8 \;. 
\ee
It allows an explicit expression in terms of $\Gamma$-functions which
however is not of particular interest for the following. Observe that
\Ref{f1} and \Ref{f2} already imply the relations
\ben
f(w)f(-w) = 1\;,\qquad f(w)^*=f(-w^*)\;.
\een
The free parameter $\a$ which appears in the solution of the QYBE is
basically a consequence of the fact that the singlet in \Ref{sodec}
may be rescaled with an arbitrary factor; two $R$ matrices \Ref{R}
with different values of $\a$ are related by conjugation with
$\mbox{diag}(\a_1\a_2^{-1},\ID_{120},\ID_{128})\otimes
\mbox{diag}(\a_1\a_2^{-1},\ID_{120},\ID_{128})$. Without loss of
generality we can thus fix the parameter $\a$ to
\be\la{alpha}
60\a^2:=-1\;.
\ee
For this value only, the $R$-matrix obeys the additional 
non-covariant relation
\be
R_{\hat{a}\hat{b}}{}^{\hat{c}\hat{d}}(w) = 
R_{\hat{c}\hat{d}}{}^{\hat{a}\hat{b}}(w)\;, \la{noncov1}
\ee
which is proved by inspection and by use of the special 
(non-covariant) property $f_a{}^{bc}= -f^a{}_{bc}$ of the 
$\EE$ structure constants \Ref{fabc}.

The following further properties of the $R$-matrix are easily verified: 
\ba
\stackrel{12}{R}(w)\stackrel{21}{R}(-w) &=& \ID ,\la{uni}\\
\stackrel{12}{R}\!(w)\,^* &=& \stackrel{21}{R}\!(-w^*)\;,\la{herm}
\ea
where the second equation is only valid for imaginary $\a$, which is
compatible with our choice \Ref{alpha} above.
In the context of two-dimensional scattering theory, these relations 
express the requirements of unitarity and hermiticity of the
S-matrix, respectively. With indices written out they acquire the 
following explicit form
\ba
R_{\hat{a}\hat{b}}{}^{\hat{g}\hat{h}}(w) 
R_{\hat{h}\hat{g}}{}^{\hat{c}\hat{d}} (-w) 
 &=& \d_{\hat{a}}^{\hat{d}}\d_{\hat{b}}^{\hat{c}}\;,\la{uni2}\\
\left(R_{\hat{a}\hat{b}}{}^{\hat{c}\hat{d}}(w) \right)^*
&=& R_{\hat{b}\hat{a}}{}^{\hat{d}\hat{c}} (-w^*)\;.\la{herm2}
\ea
The occurrence of poles at $w=w_j$ and relation \Ref{uni} together
imply that $R(w)$ is non-invertible at the points $w=-w_j$.
More specifically, \Ref{uni} yields the relations
\ba
\stackrel{12}{\cR}_j \; \stackrel{21}{R}(-w_j) = 0\;.
\ea

{}From the formulae given above it is straightforward to check that
the residue $\cR_4$ at $w_4 = 15\i$ is singled out by its property
of being proportional to a one-dimensional projector:
\be\la{R15}
\left( \cR_4 \right)_{\hat{a}\hat{b}}{}^{\hat{c}\hat{d}} = \ft{10}{7}\, 
\eta{}_{\hat{a}\hat{b}}\eta{}^{\hat{c}\hat{d}} \;,
\ee
where $\eta{}_{\hat{a}\hat{b}}$ denotes the natural extension of the
Cartan-Killing form into ${\bf 249}\otimes{\bf 249}$ given by the
additional entry $\eta{}_{00}=60\a^2=-1$. Evaluating the QYBE \Ref{QYB}
at $u\!-\!v=15\i$ then gives rise to the following relation
\be\la{r4r}
\left( \cR_4 \right)_{\hat{a}\hat{b}}{}^{\hat{g}\hat{h}}
R_{\hat{g}\hat{c}}{}^{\hat{p}\hat{i}}(u)
R_{\hat{h}\hat{i}}{}^{\hat{q}\hat{r}}(u\!-\!15\i) ~=~
\d_{\hat{c}}^{\hat{r}}\;
\left( \cR_4 \right)_{\hat{a}\hat{b}}{}^{\hat{p}\hat{q}}\;.
\ee
{}From these observations, we can deduce the crossing invariance
property of the $R$-matrix:
\be 
R_{\hat b}{}^{\hat{a}\hat{d}}{}_{\hat c} (w) \equiv
\eta^{\hat{a}\hat{g}} R_{\hat{b}\hat{g}}{}^{\hat{d}\hat{h}}(w)
\eta_{\hat{h}\hat{c}}
= R_{\hat{c}\hat{b}}{}^{\hat{a}\hat{d}}(15\i -ww)\;.\la{ci}
\ee
\bigskip

The knowledge of the $R$-matrix associated to an irreducible
representation of \Ref{yangian} now gives rise to another equivalent
presentation of the Yangian algebra itself \cite{Drin85}. Consider the
associative algebra with generators
$\left(T_{{\scriptscriptstyle(}n{\scriptscriptstyle)}}\right)
_{\hat{a}}{}^{\hat{b}}$, $(0\le \hat{a},\hat{b}\le 248)$, $n\in\N$ and
defining relations
\be
R_{\hat{a}\hat{b}}{}^{\hat{e}\hat{f}}(u-v) T_{\hat{e}}{}^{\hat{c}}(u)
 T_{\hat{f}}{}^{\hat{d}}(v) ~=~
 T_{\hat{b}}{}^{\hat{f}}(v) T_{\hat{a}}{}^{\hat{e}}(u)
 R_{\hat{e}\hat{f}}{}^{\hat{c}\hat{d}}(u-v)\la{RTT}
\ee
where $T_{\hat{a}}{}^{\hat{b}} (u)$ denotes the formal series
\be
T_{\hat{a}}{}^{\hat{b}} (u)~=~ \d_{\hat{a}}^{\hat{b}} ~+~
\sum_{n=1}^\8 
\left(T_{{\scriptscriptstyle(}n{\scriptscriptstyle)}}\right)
_{\hat{a}}{}^{\hat{b}}\;u^{-n}\;.
\ee
The QYBE \Ref{QYB} ensures compatibility of the exchange relations
\Ref{RTT} with associativity of the multiplication. Their 
evaluation at $u-v=15\i$ shows that there exists an
invariant scalar quantity $q (T(u))$, the ``quantum determinant'',
which is bilinear in the matrix entries of $T$:
\be\la{qdet0}
q (T(u)) \cR_4 ~:=~
\cR_4 \stackrel1{T}\!(u\!+\!15\i)\stackrel2{T}\!(u) ~=~
          \stackrel2{T}\!(u)\stackrel1{T}\!(u\!+\!15\i)\;\cR_4\;.
\ee
Using \Ref{r4r} one checks that $q (T(u))$ lies in the center of the
algebra \Ref{RTT}. Thus, we may pass to the quotient of this algebra
over the two-sided ideal generated by the central element by setting
$q(T)=1$ or equivalently 
\be\la{qdet}
T_{\hat{a}}{}^{\hat{c}}(u-15\i) T_{\hat{b}}{}^{\hat{d}} (u) 
\eta_{\hat{c}\hat{d}} ~=~ \eta_{\hat{a}\hat{b}}\;.
\ee
It has been stated by Drinfeld \ci{Drin85} that this quotient is
isomorphic to the Yangian $Y(\eee)$ as defined at the beginning of
this section \Ref{yangian}.\footnote{However, we presently cannot
exclude the possibility that the center of \Ref{RTT} contains elements 
of higher degree in the $T$'s which are not generated by $q(T)$.} 
The precise isomorphism requires knowledge of the universal $R$-matrix
of $Y(\eee)$ which is certainly beyond our scope here; one may however
easily identify the generating elements $\cX^a$ and $\cY^a$
\ba
\cX^a &=& \tr 
\left[X^aT_{{\scriptscriptstyle(}1{\scriptscriptstyle)}}\right]\;,\\
\cY^a &=& \tr
\left[X^a\left(T_{{\scriptscriptstyle(}2{\scriptscriptstyle)}}{}-\ft12
T_{{\scriptscriptstyle(}1{\scriptscriptstyle)}}
T_{{\scriptscriptstyle(}1{\scriptscriptstyle)}}\right)\right]
\;.\nn
\ea
We can further expand \Ref{RTT} around $u\!=\!\8$ and use \Ref{Rcl} to
obtain the commutator
\be
\Big[\stackrel{1}{T}\!\!_{{\scriptscriptstyle(}1{\scriptscriptstyle)}}
\,,\,
\stackrel{2}{T}\!(w)\Big] ~=~ 
\i\Big[\Om_\eee\,,\,\stackrel{2}{T}\!(w)\Big] \; ,\la{T1comm}
\ee
which in particular reproduces the first two commutation relations of
\Ref{yangian}. 
We close the general discussion here with two well-known properties of
the presentation \Ref{RTT} of the Yangian
\begin{itemize}
\item Any representation $\rho$ of $Y(\eee)$ defines a one-parameter
family of representations $\rho_a$ labeled by a complex number $a$: 
\be
\rho_a(T(w)) := \rho(T(w\!-\!a))\;.
\ee
The fundamental representation ${\bf 249}$ in particular gives rise to
the family ${\bf 249}_a$:  
\be\la{Va}
\r^{\bf 249}_a(T(w)) := R(w\!-\!a)\;,
\ee
with the $R$-matrix from \Ref{R}. Note that the relation \Ref{qdet}
in these representations corresponds to the normalization \Ref{r4r} of
the $R$-matrix. 

\item
The coproduct of $Y(\eee)$ takes the simple form:
\be\la{coprod}
\Delta\left(T_{\hat{a}}{}^{\hat{b}} (w)\right) = 
T_{\hat{a}}{}^{\hat{c}} (w) \otimes T_{\hat{c}}{}^{\hat{b}} (w) \;.
\ee

\end{itemize}

\mathversion{bold}
\section{Classical Yangian symmetries in $N\!=\!16$ supergravity}
\mathversion{normal} 

This section is a brief review of the classical symmetries and the
algebra of nonlocal charges in two-dimensional $N\!=\!16$ supergravity
\cite{KorSam98,NicSam98}.  The scalar sector of this model is
described by an $\EE$-valued matrix $\cV$ which transforms under a
global $\E$ symmetry and a local $SO(16)$ gauge symmetry in the usual
way
\be
\cV(x) \mapsto g\cV(x) h(x)\;,\qquad g\in\EE\;,~~h(x)\in\SO\;.  
\ee
Thus, its bosonic configuration space is given by the coset space 
$\EE/\SO$. It may be parametrized by the symmetric $\E$-valued matrix
\be\la{M}
\cM \equiv \cV\cV^T\;,\quad\mbox{i.e.~~~} 
\cM_{ab} \equiv \cV_a{}^c\cV_b{}^c = \cM_{ba}\;,
\ee
which is evidently gauge ($=SO(16)$) invariant. The symmetry of $\cM$
may be characterized algebraically by the fact that it is annihilated
by the antisymmetric projectors 
\be\la{asymM}
(\cP_{248})_{ab}{}^{cd} \cM_{cd} ~=~ 0 ~=~ 
(\cP_{30380})_{ab}{}^{cd} \cM_{cd} \;,
\ee
whereas for the symmetric projectors from \Ref{projind} one finds the
following identities 
\ba
(\cP_{1})_{ab}{}^{cd} \cM_{cd} &=& \ft1{31} \eta{}_{ab} 
\quad  \Longleftrightarrow \quad \eta^{ab} \cM_{ab} = 8 \;,\la{symM}\\
(\cP_{3875})_{ab}{}^{cd} \cM_{cd} &=& 0 \;,  \non
(\cP_{27000})_{ab}{}^{cd} \cM_{cd} &=& \cM_{ab} - \ft1{31} 
\eta{}_{ab}\;.\nn 
\ea
To verify these relations one needs the $\E$-invariance of the
projectors and the Cartan-Killing form: 
\ben
(\cP_{j})_{ab}{}^{cd} M_{cd} ~=~ \cV_a{}^{c} \cV_b{}^{d}
(\cP_{j})_{cd}{}^{ee} \;,\qquad
\cV_a{}^{c} \cV_b{}^{d}\:\eta_{cd}~=~\eta_{ab}\;.
\een
The second relation in \Ref{symM} requires the additional formula
\[ f^d{}_{ac} f_{dbc} =
   \left\{ \begin{array}{ll}
           8\delta^{IJ}_{KL}  & \mbox{if $(ab)=(IJ,KL)$} \\
           0 & \mbox{otherwise}
           \end{array}
           \right.    
\]
where the summation over the $\E$ index $c$ is with the ``wrong
metric'' (i.e. with the $\SO$-covariant $\delta_{ab}$ rather than
$\eta_{ab}$).
\bigskip

The scalar fields represented by the matrix $\cM$ satisfy equations of
motion which allow a Lax pair formulation
\ci{BelZak78,Mais78,BreMai87,Nico87,NicWar89} similar to the principal
chiral model. In particular this allows the construction of an
infinite family of nonlocal integrals of motion which are obtained
from the transition matrices associated to the Lax pair
\ci{KorSam98,NicSam98}. These integrals of motion are encoded in a
symmetric $\E$-valued matrix $\cM(w)$ obtained by integrating the Lax
connection over certain space intervals and depending on a complex
spectral parameter $w$. This matrix parametrizes the full scalar
sector of the phase space in the sense that for real values of $w$ the
matrix $\cM(w)$ coincides with the physical scalar fields $\cM(x)$
evaluated on the particular axis in space-time where the dilaton field
$\r$ vanishes
\be
\cM(w) = \cM(x)\Big|_{\r(x)=0,\: \rt(x)=w}\;\;.
\ee
This relation has been formulated in the coordinate system where the
two-dimensional world-sheet is parametrized by the dilaton field $\r$
and its dual axion $\rt$. For instance, for cylindrically symmetry 
spacetimes the matrix $\cM(w)$ carries the values of physical scalar 
fields along the symmetry axis. 

We may further introduce its Riemann-Hilbert decomposition 
\be\la{Upm}
\cM_{ab}(w) \equiv U_+(w)_a{}^c \,U_-(w)_b{}^c\;,
\ee
into $E_8$-valued functions $U_\pm (w)$ which are holomorphic in the
upper and the lower half of the complex $w$-plane, respectively.  
They are related by complex conjugation
\be\la{coco}
\left(U_{\!+}(w)\right)^* ~=~ U_{\!-}(w^*)\;.
\ee

In \ci{KorSam98,NicSam98} it was shown that these phase space
quantities are subject to the following symplectic structure:
\be\la{PBM}
\left\{\cM_{ab}(v)\;,\,\cM_{cd}(w)\right\} ~~=
\ee
\vspace*{-1em}
\ba
&&  \frac1{v-w}\Big((\Om_\eee)_{ac}{}^{mn}\cM_{mb}(v)\cM_{nd}(w)\,
+  \cM_{am}(v)\cM_{cn}(w) \, (\Om_\eee)_{bd}{}^{mn} \non
&& {} -
\cM_{am}(v) (\Om_\eee^\t)_{mc}{}^{bn}\cM_{nd}(w)
\,- \cM_{cm}(w)\,(\Om_\eee^\t)_{an}{}^{md}\cM_{nb}(v)\Big) 
\;,\nn
\ea
with $\Om_\eee$ and $\Om_\eee^\t$ from \Ref{O1} and \Ref{O2},
respectively. One may check that these Poisson brackets are covariant
under $\E$ and compatible with the symmetry of $\cM$ \Ref{asymM}, as
required for consistency. For the purpose of quantization to be
addressed in the next section it is further convenient to decompose
this structure according to \Ref{Upm} into the following brackets
\ba
\left\{\stackrel1{U_\pm }\!(v)\;,\,\stackrel2{U_\pm }\!(w)\right\}
&=& \left[\frac{2\Om_\eee}{v-w}\,,\, 
\stackrel1{U_\pm }\!(v)\stackrel2{U_\pm }\!(w)
\right]\;,\la{U1}\\
\left\{\stackrel1{U_\pm }\!(v)\;,\,\stackrel2{U_\mp}\!(w)\right\}
&=& \frac{2\Om_\eee}{v-w} \stackrel1{U_\pm }\!(v)\stackrel2{U_\mp}\!(w)
-\stackrel1{U_\pm }\!(v)
\stackrel2{U_\mp}\!(w)\frac{2\Om_\eee^\t}{v-w}\;.
\nn\ea

In a theory with local symmetries, observables such as the conserved
non-local charges contained in $U_\pm (w)$ must weakly commute with the
associated canonical constraints. For the above charges this was shown
to be the case in \cite{NicSam98}. Namely, for the traceless
components $T'_{\mu \nu}:= T_{\mu \nu} -\ft12 g_{\mu \nu} T^{\rho}{}_\rho$ of
the energy momentum tensor (generating local translations along the
lightcone), we simply have
\be
\Big\{ T'_{\mu \nu} (z) \, , \, U_\pm (w) \Big\} ~=~ 0\;.   \la{UT}
\ee
This relation expresses the invariance of the charges $U_\pm (w)$ 
under general coordinate transformations, which thus indeed
constitute ``observables'' in the sense of Dirac.

In supergravity, we have in addition the constraints $S^I_\alpha (z)$
generating $N\!=\!16$ local supersymmetry transformations ($\alpha$ is
a spinor index in two dimensions). As shown in \cite{NicSam98}, the
relations expressing the invariance of the charges $U_\pm (w)$ under
local supersymmetry are considerably more complicated than \Ref{UT}.
Recalling that the integrals of motion $U_\pm (w)$ are obtained from
certain transition matrices $U(x,y;w)$ associated to the Lax pair of
the model, we found that they obey Poisson bracket of the following
type (for $x<z<y$):
\be
\Big\{U(x,y;w),S^I_\alpha (z)\Big\} ~~\sim~~
U(x,z;w)\,X^{IJ}S^J_\alpha (z)\,U(z,y;w)\;,\la{US}
\ee
which vanish indeed on the constraint surface $S^I_\alpha (x) =0$.
Apart from questions of operator ordering, it is clear from the form
of \Ref{US} that the combined algebra of nonlocal charges and
supersymmetry constraints does not close. It remains an open problem
at this point whether one can arrive at a closed structure upon
sufficient enlargement of the algebra. Its quantization would entail
the existence of a novel type of exchange relations between the
conserved charges and the local supersymmetry constraints. The full
algebra should then contain the Yangian charge algebra to be presented
in the next section as well as a quantized version of the $N\!=\!16$
superconformal algebra, into which the supersymmetry constraints
close. Note however, that a consistent quantum formulation of the
latter is a highly nontrivial task due to the nonlinear nature of the
$N\!=\!16$ superconformal algebra. For instance, -- and in contrast to
the standard extended superconformal algebras -- free field
realizations are not even known at the classical level.

\section{Quantization}
We now wish to quantize the symplectic structure of the classical
charge algebra by means of the $R$-matrix described above. This
amounts to replacing the Poisson brackets \Ref{U1} by quantum exchange
relations, leading to a ``twisted'' Yangian double with central
extension $c$.  More precisely, we employ the construction \Ref{RTT}
to replace the classically conserved non-local charges $U_\pm (w)$ (which
by their definition are $248\times 248$ matrices) by a corresponding set of
$249\times 249$ matrices $T_\pm (w)$ with operator-valued entries subject to
the exchange relations
\ba
\stackrel{12}{R}\!(v\!-\!w)\stackrel1{T_\pm }\!(v)\stackrel2{T_\pm }\!(w) &=&
\stackrel2{T_\pm }\!(w)\stackrel1{T_\pm }\!(v) \;\stackrel{12}{R}\!(v\!-\!w)
\;,\la{YN1}\\
\stackrel{12}{R}\!(v\!-\!w\!-\! \i c)
\stackrel1{T_-}\!(v)\stackrel2{T_+}\!(w) &=&
\stackrel2{T_+}\!(w)\stackrel1{T_-}\!(v) \;
\stackrel{12}{Q}\!(v\!-\!w)\la{YN2}\;,
\ea
where for the ``twisted'' $R$-matrix $Q$ we require the classical
expansion (cf. \Ref{Rcl})
\ba
Q(w) = \ID - \frac{\i}{w} \Omega_\eee^\t  \, + \, 
{\cal O} \left(\frac{\hbar^2}{w^2}\right) \;,\la{Q}
\ea
and the compatibility relations
\ba
\stackrel{12}Q\!(u\!-\!v) \stackrel{13} Q\!(u) \stackrel{23} R\!(v) 
&~=~&
\stackrel{23}R\!(v) \stackrel{13} Q\!(u) \stackrel{12} Q\!(u\!-\!v) 
\;,\la{RQQ}\\
\stackrel{12}R\!(u\!-\!v) \stackrel{13} Q\!(u) \stackrel{23} Q\!(v) 
&~=~&
\stackrel{23}Q\!(v) \stackrel{13} Q\!(u) \stackrel{12} R\!(u\!-\!v) 
\;, \nn
\ea
whose derivation is completely analogous to \Ref{QYB}. For
$60\alpha^2=-1$,  the solution is given by
\be\la{QR}
Q_{\hat{a} \hat{b}}{}^{\hat{c} \hat{d}} (w) ~:=~ 
R_{\hat{d} \hat{a}}{}^{\hat{b} \hat{c}} (-w) =
R_{\hat{b} \hat{c}}{}^{\hat{d} \hat{a}} (-w)\;,
\ee
i.e. by interchanging the two subspaces and taking the transpose of the 
original $R$-matrix in one of them. The interchange of subspaces here 
is necessary because $\stackrel{12}{R}\neq \stackrel{21}{R}$.
It is easy to check that the above definition yields the correct first 
order term displayed in \Ref{Q}. Furthermore, although transposing 
the indices is a non-covariant operation, it turns out that all 
summations in \Ref{RQQ} are again covariant, such that with a little 
algebra these relations can be reduced to the original QYBE \Ref{QYB}. 

We emphasize that the shift $c$ (alias the central charge) in \Ref{YN2} 
is compatible with all of our requirements so far and therefore still 
arbitrary at this point. It is important here that the algebras for
different $c$ are not isomorphic; in particular, they may have
different ideals. The central charge $c$ will be fixed later by
requiring symmetry of the quantum monodromy matrix. Note that a
possible additional shift in the argument of $Q$ in \Ref{QR} has been
absorbed into a redefinition of $T_-$.

As for the singular points, there is an important difference between
\Ref{YN1} and \Ref{YN2}: whereas the poles on the l.h.s.~and r.h.s.~of
\Ref{YN1} always match, this is not so for \Ref{YN2} due to the
shift. Thus, either some of the mixed operator products are singular,
or the regularity on one side imposes the vanishing of certain
residues on the other side. These questions as well as the proper
quantum analogue of the classical holomorphy properties of the $T_\pm $'s
may however only be addressed after specializing to a particular
representation of \Ref{YN1}, \Ref{YN2}. To be on the safe side here,
we will use the exchange relations only at the generic points where
the $R$-matrices are nonsingular.

In addition to these exchange relations we demand that the quantum
determinant for both $T_+$ and $T_-$ be equal to unity,
viz.~\Ref{qdet} 
\be\la{qdetTpm}
T_\pm (w-15\i)_{\hat{a}}{}^{\hat{c}}\: T_\pm (w)_{\hat{b}}{}^{\hat{d}} 
\;\eta_{\hat{c}\hat{d}} ~=~ \eta_{\hat{a}\hat{b}}\;.
\ee
Due to \Ref{f1}, $Q$ in \Ref{QR} is normalized such that the l.h.s.~of
this equation indeed lies in the center of the full algebra \Ref{YN1},
\Ref{YN2}.  The hermiticity \Ref{herm} of the $R$-matrix shows that
the full quantum algebra is compatible with the following
$*$-structure -- suggested by the classical relation \Ref{coco} --
\be\label{*}
\left(T_\pm (w)_{\hat{a}}{}^{\hat{b}}\right)^* = 
T_\mp(w^*)_{\hat{a}}{}^{\hat{b}} \;,
\ee
for the purely imaginary choice of the parameter $\alpha$ we have
made.\footnote{It is helpful to note that like for $\gg =\mathfrak{sl}_2$
\cite{KorSam98} the algebra \Ref{YN1}, \Ref{YN2} may in fact be mapped
to the usual (untwisted) centrally extended Yangian double ${\cal
D}Y(\eee)_c$ by the (noncovariant) map
$$T_+(w)_{\hat{a}}{}^{\hat{b}}\mapsto
T_+(w)_{\hat{a}}{}^{\hat{c}}\eta_{{\hat{c}}{\hat{b}}} \;,\quad
T_-(w)_{\hat{a}}{}^{\hat{b}}\mapsto T_-(w)_{\hat{a}}{}^{\hat{b}}\;.$$
The additional relation \Ref{noncov1} is required to show that this is
indeed an automorphism of \Ref{YN1}.
With respect to \Ref{*} this map is, however, no $*$-isomorphism; the
representation theory of \Ref{YN1}, \Ref{YN2} will thus differ
considerably from the one of ${\cal{D}}Y(\eee)_c$. (Needless to say
that even the latter is far from being developed.)}

Let us show that the algebra \Ref{YN1}--\Ref{*} has the correct
classical limit $\hbar\ra0$. If we embed the original non-local
charges $U_\pm (w)$ by identifying them with the upper left $248\times 248$
block of $T_\pm (w)$, the exchange relations \Ref{YN1}, \Ref{YN2} reduce
to the Poisson brackets \Ref{U1} in the limit $\hbar\rightarrow
0$. The $U_\pm (w)$ being $\E$-valued matrices, the condition
\Ref{qdetTpm} can then be viewed as the quantum analog of the
statement that any element of $E_{8(8)}$ also belongs to
$SO(128,120)$. While this submatrix of $T_\pm $ is evidently the quantum
analog of the classical charges, one may wonder about the significance
of the extra components $T_0{}^a(w), T_a{}^0(w)$ and the singlet
$T_0{}^0(w)$. The exchange relations may be read in such a way, that
the off-diagonal components can be solved to become functions of the
248 degrees of freedom originally present. In order to make the
dependence explicit, we evaluate the defining relation \Ref{RTT} at
the remaining poles $u-v=w_j$ $(j=1,\dots,3)$ (the residue at $w_4$
has already been exploited to derive \Ref{qdet}):
\ben
\stackrel{12}{\cR}_j \,\stackrel{1}{T}\!(u)\stackrel{2}{T}\!(u\!-\!w_j)~=~
\stackrel{2}{T}\!(u\!-\!w_j) \stackrel{1}{T}\!(u) \stackrel{12}{\cR}_j\,. 
\een
After expansion around $u=\8$ this equation can be solved order by
order to get expressions for the off-diagonal components
$(T_{{\scriptscriptstyle(}n{\scriptscriptstyle)}})_0{}^a$ and
$(T_{{\scriptscriptstyle(}n{\scriptscriptstyle)}})_a{}^0$,
respectively. In first order this yields
\ben
\stackrel{12}{\cP}_j \;
\stackrel{1}{T}\!\!_{{\scriptscriptstyle(}1{\scriptscriptstyle)}}
+\stackrel{2}{T}\!\!_{{\scriptscriptstyle(}1{\scriptscriptstyle)}} ~=~
\stackrel{2}{T}\!\!_{{\scriptscriptstyle(}1{\scriptscriptstyle)}}+ 
\stackrel{1}{T}\!\!_{{\scriptscriptstyle(}1{\scriptscriptstyle)}}\; 
\stackrel{12}{\cP}_j\;,
\een
for all projectors from \Ref{projind} and \Ref{proj2}. Hence, 
\be
(T_{{\scriptscriptstyle(}1{\scriptscriptstyle)}})_a{}^b \in\eee \;,
\quad \mbox{and}\;\;\;
(T_{{\scriptscriptstyle(}1{\scriptscriptstyle)}})_0{}^a =
(T_{{\scriptscriptstyle(}1{\scriptscriptstyle)}})_a{}^0 =
(T_{{\scriptscriptstyle(}1{\scriptscriptstyle)}})_0{}^0 =0\;.
\ee
In second order we get the equations
\ben\begin{array}{l}
\stackrel{12}{\cR}_j \,\Big(\!
\stackrel{1}{T}\!\!_{{\scriptscriptstyle(}2{\scriptscriptstyle)}}
 -\ft12(
\stackrel{1}{T}\!\!_{{\scriptscriptstyle(}1{\scriptscriptstyle)}})^2+ 
\stackrel{2}{T}\!\!_{{\scriptscriptstyle(}2{\scriptscriptstyle)}}
 -\ft12(
\stackrel{2}{T}\!\!_{{\scriptscriptstyle(}1{\scriptscriptstyle)}})^2 
+ w_j \stackrel{2}{T}\!\!_{{\scriptscriptstyle(}1{\scriptscriptstyle)}}
+\ft12 \Big[\!
\stackrel{1}{T}\!\!_{{\scriptscriptstyle(}1{\scriptscriptstyle)}},
\stackrel{2}{T}\!\!_{{\scriptscriptstyle(}1{\scriptscriptstyle)}}\!\Big]
\Big)\\[5pt]
\;\;=\Big(\!
\stackrel{1}{T}\!\!_{{\scriptscriptstyle(}2{\scriptscriptstyle)}} 
-\ft12(
\stackrel{1}{T}\!\!_{{\scriptscriptstyle(}1{\scriptscriptstyle)}})^2 +
\stackrel{2}{T}\!\!_{{\scriptscriptstyle(}2{\scriptscriptstyle)}} 
-\ft12(
\stackrel{2}{T}\!\!_{{\scriptscriptstyle(}1{\scriptscriptstyle)}})^2 
+ w_j \stackrel{2}{T}\!\!_{{\scriptscriptstyle(}1{\scriptscriptstyle)}} 
-\ft12 \Big[\!
\stackrel{1}{T}\!\!_{{\scriptscriptstyle(}1{\scriptscriptstyle)}},
\stackrel{2}{T}\!\!_{{\scriptscriptstyle(}1{\scriptscriptstyle)}}\!\Big]
\Big)\,\stackrel{12}{\cR}_j\;.\end{array}
\een
Together with \Ref{T1comm}, \Ref{qdet} and the explicit form of the
residues of the $R$-matrix \Ref{resi} these relations can be used to
deduce 
\ba
&&\Big(
T_{{\scriptscriptstyle(}2{\scriptscriptstyle)}} 
-\ft12 T_{{\scriptscriptstyle(}1{\scriptscriptstyle)}}
T_{{\scriptscriptstyle(}1{\scriptscriptstyle)}}\Big)_a{}^b~
\in~\eee\;,\\
&&(T_{{\scriptscriptstyle(}2{\scriptscriptstyle)}})_0{}^a ~=~ 
\ft{\sqrt2}{\a}\; \i f^{ab}{}_c\,
(T_{{\scriptscriptstyle(}1{\scriptscriptstyle)}})_c{}^b  
\;,\quad
(T_{{\scriptscriptstyle(}2{\scriptscriptstyle)}})_0{}^0 ~=~ 0\;,
\;\;\mbox{etc.} \nn
\ea
In this fashion one may in principle determine the components
$T(w)_0{}^a, T(w)_a{}^0$ in all orders as functions of the
$T(w)_a{}^b$ which vanish in the classical limit $\hbar\rightarrow
0$. Thus, we have consistently
\be\la{class}
T_\pm (w)_{\hat{a}}{}^{\hat{b}}\; 
\begin{tabular}{c}\\[-5pt] $\hbox to 1.1cm {\rightarrowfill}$\\[-5pt]
$\scriptstyle{\hbar\ra 0}$\end{tabular}\;
\left(\!\!\mbox{\begin{tabular}{c|c} 
$U_\pm (w)_a{}^b$&$0$\\ \hline $0$&$1$\end{tabular}}\!\!\right) \;.
\ee
\bigskip

Recall now that the classical phase space was parametrized by the
symmetric $\E$-valued matrix $\cM_{ab}(w)$. On the quantum side we
define this object in analogy to \Ref{Upm} as
\be\la{Upmq}
\cM_{\hat{a}\hat{b}}(w) \equiv T_+(w)_{\hat{a}}{}^{\hat{c}}
\,T_-(w)_{\hat{b}}{}^{\hat{c}}\;, 
\ee
where the operator ordering on the r.h.s. is fixed by this
relation. The matrix entries of $\cM$ are distinguished elements in
the algebra \Ref{YN1}, \Ref{YN2} in that they satify the exchange
relations  
\ba
T_+(v)_{\hat{c}}{}^{\hat{d}}\, \cM_{\hat{a}\hat{b}}(w) &\!=\!&
R_{\hat{a} \hat{c}}{}^{\hat{p} 
\hat{k}}(w\!-\!v) \cM_{\hat{p}\hat{q}}(w)
R_{\hat{b} \hat{k}}{}^{\hat{q} \hat{l}}(w\!-\!v\!-\!\i c)
\,T_+(v)_{\hat{l}}{}^{\hat{d}} \;,\non
T_-(v)_{\hat{c}}{}^{\hat{d}}\, \cM_{\hat{a}\hat{b}}(w) &\!=\!&
R_{\hat{a} \hat{c}}{}^{\hat{p} \hat{k}}(w\!-\!v\!+\!\i c) 
\cM_{\hat{p}\hat{q}}(w)
R_{\hat{b} \hat{k}}{}^{\hat{q} \hat{l}}(w\!-\!v)
\,T_-(v)_{\hat{l}}{}^{\hat{d}} \;,\non[4pt]
&&\la{TMR}
\ea
as well as the closed algebra
\be\la{MM}
R_{\hat{a} \hat{b}}{}^{\hat{m} \hat{n}}(v\!-\!w)
\;\cM_{\hat{m}\hat{k}}(v)
\;R_{\hat{c} \hat{n}}{}^{\hat{k} \hat{l}}(v\!-\!w\!-\!\i c) 
\;\cM_{\hat{l}\hat{d}}(w)~=~
\ee
\vspace*{-1em}
\ben
\hspace*{8em}
\cM_{\hat{b}\hat{m}}(w)
\;R_{\hat{k} \hat{a}}{}^{\hat{m} \hat{n}}(w\!-\!v\!-\!\i c)
\;\cM_{\hat{n}\hat{l}}(v)
\;R_{\hat{d} \hat{c}}{}^{\hat{k} \hat{l}}(w\!-\!v)\;,
\een
which we hence view as the quantized version of
\Ref{PBM}.\footnote{Like its classical counterpart \Ref{PBM} the
algebra \Ref{MM} belongs to the general class of quadratic algebras
which has been considered in \cite{FreMai91}.}

While the classical matrix $\cM$ was manifestly symmetric
(cf. \Ref{asymM}, \Ref{symM}) this is not necessarily true for its
quantum analog.  Rather, we must now impose some quantum version of
this condition in order to ensure that the number of degrees of
freedom in the quantum structure matches the classical phase space. In
other words, we still have to implement the quantum analogue of the
classical coset structure. In algebraic language this amounts to
dividing out another ideal from \Ref{YN1}--\Ref{YN2}. Unlike the
quantum determinant condition \Ref{qdetTpm} (which we still assume to
hold), this new condition will involve $T_+$ and $T_-$ simultaneously.

To this end we return to the exchange algebra \Ref{YN1}--\Ref{YN2}
with arbitrary, but fixed central charge $c$, and consider the set of
elements
\be\la{ideal}
\phi_{\hat{a} \hat{b}}(w) ~\equiv~ \left\{
   \begin{array}{ll} 
       \mbox{Res}\Big|_{v=\i c}
R_{\hat{a}\hat{b}}{}^{\hat{c}\hat{d}}(v)\;\cM_{\hat{c}\hat{d}}(w)&
         \;\;\mbox{if $R(\i c)$ is singular}\\[12pt]
R_{\hat{a}\hat{b}}{}^{\hat{c}\hat{d}}(\i c)\cM_{\hat{c}\hat{d}}(w)&  
            \;\;\mbox{else}\;.
                  \end{array}\right.
\ee
Use of the exchange relations \Ref{TMR} then yields in a first step
\ben
T_+(u)_{\hat{c}}{}^{\hat{d}} \;\phi_{\hat{a} \hat{b}}(w) ~=~
\een
\vspace*{-1.2em}
\ba\;\;&&~
  =~R_{\hat{a} \hat{b}}{}^{\hat{p} \hat{q}}(\i c)
R_{\hat{p} \hat{c}}{}^{\hat{m} \hat{k}}(w\!-\!u) 
R_{\hat{q} \hat{k}}{}^{\hat{n} \hat{l}}(w\!-\!u\!-\!\i c)
\;\cM_{\hat{m}\hat{n}}(w)\,T_+(u)_{\hat{l}}{}^{\hat{d}} \;.\nn
\ea
The necessity of the choice $v=\i c$ in \Ref{ideal} becomes evident 
at this point: it is the only value of the argument of the $R$-matrix
in \Ref{ideal} for which we can exploit the QYBE to re-arrange 
the indices (this would not be possible if the first factor on the 
r.h.s. were $R_{\hat{a} \hat{b}}{}^{\hat{p} \hat{q}}(v)$ with arbitrary 
argument $v$). Thus, by use of \Ref{QYB} we finally obtain
\ba\;\;&&~=~ 
R_{\hat{b} \hat{c}}{}^{\hat{q} \hat{k}}(w\!-\!u\!-\!\i c)
R_{\hat{a} \hat{k}}{}^{\hat{p} \hat{l}}(w\!-\!u) 
R_{\hat{p} \hat{q}}{}^{\hat{m} \hat{n}}(\i c)
\;\cM_{\hat{m}\hat{n}}(w)\,T_+(u)_{\hat{l}}{}^{\hat{d}}  \non
&&~=~
R_{\hat{b} \hat{c}}{}^{\hat{q} \hat{k}}(w\!-\!u\!-\!\i c)
R_{\hat{a} \hat{k}}{}^{\hat{p} \hat{l}}(w\!-\!u) 
\;\phi_{\hat{p} \hat{q}}(w)\,T_+(u)_{\hat{l}}{}^{\hat{d}}\;. \la{clc}
\ea
A similar calculation for $T_-(u)$ gives the same result. We conclude
that the elements $\phi_{\hat{a} \hat{b}}(w)$ constitute the basis of
a two-sided ideal of \Ref{YN1}--\Ref{YN2}. Obviously, these ideals are
nontrivial only if \Ref{ideal} does not contain $\cM$ entirely,
i.e.~only if $R(\i c)$ or the relevant residue is singular or
non-invertible. This happens only at the special values
$c=\pm w_j$.\footnote{Together with the fact \Ref{R15} that $\cR_4$ is
proportional to a one-dimensional projector, \Ref{clc} in particular
shows the well-known infinite-dimensional enlargement of the center of
the algebra \Ref{YN1}--\Ref{YN2} at the critical level $c=15$
\ci{ResSem90}. However, to achieve consistency with the classical
coset structure $\E/\SO$ we need another value of the central
extension here.}

There is (of course) a more group-theoretical interpretation of this
construction: in view of \Ref{Va} and \Ref{coprod}, the exchange
relations \Ref{TMR} express the fact, that under the adjoint action of
$T_+(v)$ and $T_-(v)$, respectively, the matrix $\cM(w)$ transforms in
the tensor product ${\bf 249}_{w}\otimes{\bf 249}_{w-\i c}$. I.e.~the
existence of nontrivial ideals in $\cM$ amounts to the reducibility of
the tensor product ${\bf 249}_{\i c}\otimes{\bf 249}$ which is in
correspondence with the singular points of the associated $R$-matrix
\cite{ChaPre91a} as we have explicitly seen here.

Returning to the problem of identifying the proper quantum analogue of
the symmetry of $\cM$, let us now examine \Ref{ideal} for all critical
choices of the central extension $c$ with the desired conditions
\Ref{asymM}, \Ref{symM}. With the explicit form of \Ref{resi} one
confirms that there is a unique value of $c$ such that the algebra
\Ref{YN1}--\Ref{YN2} has an ideal which in the classical limit indeed
reduces to \Ref{asymM} and \Ref{symM}. The correct choice is
\be
c=1\;.
\ee
Dividing out the ideal corresponding to \Ref{ideal} now amounts to 
imposing the additional set of relations $\phi_{\hat{a}\hat{b}}=0$, or
\be
\left(\cR_1\right)_{\hat{a}\hat{b}}{}^{\hat{c}\hat{d}} 
\cM_{\hat{c}\hat{d}}(w) ~=~ 0 \;.
\ee
The ensuing relations can be written more succinctly by splitting
$\cR_1$ into the $\E$-invariant projectors: 
\ba
(\cP_{248})_{ab}{}^{cd} \cM_{cd} &=& 
- \ft{\a}8\,f_{ab}{}^c(\cM_{0c}+\cM_{c0})\;, \la{iderel}\\
(\cP_{3875})_{ab}{}^{cd} \cM_{cd} &=& 0 \;,\non
(\cP_{30380})_{ab}{}^{cd} \cM_{cd} &=& 0 \;,\non
\cM_{0c} &=& \cM_{c0} \;,\non
\cM_{00} &=& -\ft1{480\a^2}\,\eta^{ab}\cM_{ab} \;.\nn
\ea
Since the off-diagonal components $\cM_{0a}$ are of order
${\cal O}(\hbar)$ by \Ref{class}, it now follows with our choice
$60\a^2=-1$ that the relations \Ref{iderel} indeed encompass 
the classical coset relations \Ref{asymM} and \Ref{symM} in the 
limit $\hbar\rightarrow 0$.

In conclusion, the quantum algebra which replaces the classical
Poisson algebra \Ref{U1} is given by \Ref{YN1}--\Ref{*} with central
extension $c\!=\!1$ divided by the ideal which is generated by
\Ref{iderel}. The operator \Ref{Upmq} consistently represents the
quantum analogue of the classical matrix $\cM(w)$ related to the
physical scalar fields on a certain axis in space-time. Matrix
elements of \Ref{Upmq} in particular representations should thus carry
the information about quantum spectra and fluctuations of the original
fields. These issues as well as the general representation theory of
the twisted Yangian doubles remain to be investigated.

\paragraph{Acknowledgements} We thank D.~Bernard, B.~Julia,
J.M.~Maillet and M.~Niedermaier for helpful comments and
conversations. This work was supported by EU contract
ERBFMRX-CT96-0012.

\providecommand{\href}[2]{#2}\begingroup\raggedright
\endgroup

\end{document}